\documentclass[twocolumn,showpacs,amssymb,preprintnumbers,prc]{revtex4} 
\usepackage{rotating,epsfig} \usepackage{graphicx}

\begin{document}
\title{Mirror displacement energies and neutron skins.}
\author{ J. Duflo$^a$, and A. P. ~Zuker$^b$ }
\affiliation{
(a) Centre de Spectrom\'etrie Nucl\'eaire et de Spectrom\'etrie
de Masse (IN2P3-CNRS) 91405 Orsay Campus, France\\
(b) IReS, B\^at27, IN2P3-CNRS/Universit\'e Louis
Pasteur BP 28, F-67037 Strasbourg Cedex 2, France}
\date{\today}
\begin{abstract}
  A gross estimate of the neutron skin [0.80(5)$(N-Z)/A$ fm] is
  extracted from experimental proton radii, represented by a four
  parameter fit, and observed mirror displacement energies (CDE). The
  calculation of the latter relies on an accurately derived Coulomb
  energy and smooth averages of the charge symmetry breaking
  potentials constrained to state of the art values. The only free
  parameter is the neutron skin itself.  The Nolen Schiffer anomaly is
  reduced to small deviations (rms=127 keV) that exhibit a secular
  trend. It is argued that with state of the art shell model
  calculations the anomaly should disappear. Highly accurate fits to
  proton radii emerge as a fringe benefit.
\end{abstract}
\maketitle
\begin{figure}[h]
  \begin{center}
    \leavevmode
    \epsfig{file=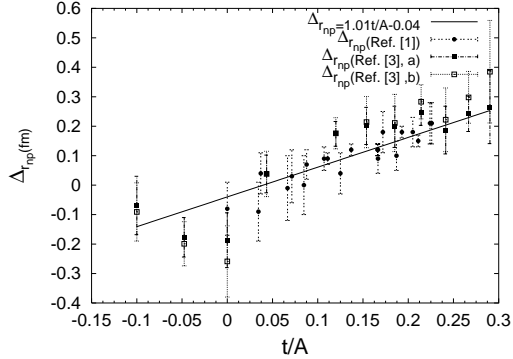,width=7cm}
    \caption{$\Delta_{r_{\nu \pi}}$ from:~[1, Fig.~4]
      (circles), the two approximations of Ref.~[3, Table
      I] (squares), and Eq.~(\ref{eq:s1})}
    \label{fig:skins}
  \end{center}
\end{figure}

Recent experiments~\cite{trz01,kra99} have considerably added to our
knowledge of neutron radii, the most elusive of the fundamental
properties of nuclear ground states. The two sets of measures are
consistent with one another, and a recognizable pattern
emerges~\cite[Fig.~4]{trz01}, leading to an estimate for the neutron
skin ($\nu\equiv$ neutrons, $\pi\equiv$ protons, $t=N-Z$)
\begin{equation}
  \label{eq:s1}
  \Delta_{r_{\nu \pi}}=
  \sqrt{\langle r_{\nu}^2\rangle}-\sqrt{\langle
  r_{\pi}^2\rangle}=-0.04(3)+1.01(15)\frac{t}{A}\, \text{fm}. 
\end{equation}
A third---totally different---experiment~\cite{suz95} adds
weight to this estimate: it deals with the sodium isotopes,
lighter and far more exotic than the species studied
in~\cite{trz01,kra99}. Nonetheless, their $\Delta_{r_{\nu \pi}}$
behaviour is very much the same, as seen in Fig.~\ref{fig:skins}.

On the theory side, we have an elegant analysis of
$\Delta_{r_{\nu \pi}}$\cite{pet96}, and
many mean-field calculations. Some of them are close to
estimate~(\ref{eq:s1}), others not so
close~\cite{dob96}, though in the Na isotopes several Skyrme forces
seem to give equally good results~\cite{bro96}. As was pointed out
in~\cite{gom83}, the neutron skins could be easily varied in such
forces, without perturbing other observables, but the criteria to fix
them were not obvious. The problem has been tackled head on
recently~\cite{cha97,bro00} by constraining agreement with realistic
calculations of the equation of state for neutron matter but no
systematic study has appeared so far.

The idea that started this investigation is that---assuming isospin
conservation---a complete knowledge of {\em proton} radii would
determine $\Delta_{r_{\nu \pi}}$. As the available experimental data
turn out to be consistent with a wide range of possible values of
$\Delta_{r_{\nu \pi}}$, we resort to the Coulomb displacement energies
(CDE) {\em which are very sensitive to this quantity}. The originality
of our approach rests on the derivation, for all the observables, of
smooth---Bethe Weizs\"acker type---forms that depend only on $N$ and
$Z$.  The differences between these averages and the experimental
values---i.e., the shell effects--- will be found to exhibit secular
behaviour that leads to high quality phenomenological fits for the
radii. For the CDE, they lead to an optimistic assessment of the
present status of a famous problem, the Nolen Schiffer
anomaly~\cite{nol69}.

To relate neutron and proton radii we start by noting that
$r_{\pi}^2=\left[\sum_{i<j}(1/2-t_z^i)(1/2-t_z^j)r_{ij}^2\right]/A^2$
has the same form as the Coulomb energy (interchange $r_{ij}^2$ by
$r_{ij}^{-1}$) . It follows that $\langle r_{\pi}^2\rangle=\alpha
(A,T)+\beta (A,T)T_z+\gamma (A,T)T_z^2$. This result relies on the
same arguments that lead to the isobaric multiplet mass
equation~\cite[p.  302]{jan69}. As $\langle r_{\pi}^2\rangle$ is a
functional of the state occupancies, $\alpha,\, \beta$ and $\gamma$
may vary rapidly. To obtain a smooth form we assume some continuous
occupancies, which make it possible to obtain a four-parameter
expression based only on dimensional considerations.  Since we are
interested only in mirror nuclei with $T_z=2t$, we propose
 \begin{equation}
\label{eq:R}
 \sqrt{\langle r_{\pi}^2\rangle}\approx \rho_{\pi}=
A^{1/3}\left(\rho_0-\frac{\zeta}{2}\frac{t}{A^{\sigma}}-
\frac{\upsilon}{2}(\frac{t}{A})^2\right)e^{(g/A)}.     
\end{equation}
The overall $A^{1/3}$ dependence is a general asymptotic result for
self-bound systems.  The $\exp (g/A)$ correction accounts for the
larger radii for small $A$. The terms in $\upsilon$ and $\zeta$ must
be {\em at most} of the same order as $\rho_0$ for large $t/A$ and the
only remaining uncertainty is in the scaling $\sigma$ in
$t/A^{\sigma}$. 

{\em Obviously, we can derive a similar expression for $\rho_{\nu}$
  with $t\Rightarrow -t$}. Therefore $\upsilon>0$ represents a uniform
contraction of the two fluids, while $\zeta>0$ implies a
$\pi$-contraction and $\nu$-dilation, which gives the neutron skin:
\begin{equation}
  \label{eq:skin}
 \Delta_{r_{\nu \pi}}\equiv
 \Delta(\zeta)=(\rho_{\nu}-\rho_{\pi})=\frac{\zeta t}{A^{\sigma-1/3}}e^{(g/A)}.  
\end{equation}

If $\sigma=1$  the volumes occupied by the two fluids may differ by a
quantity of order $A$, and we have a ``volume'' skin.

If $\sigma=4/3$ we have a ``surface'' skin, since the difference in volumes
is at most of order $A^{2/3}$.

The volume option would be the correct one for strong attraction
between like particles. In nuclei, the $\nu\, \pi$ force is by far the
strongest, and we adopt a surface skin. (A volume skin
also leads to excellent radii, but it will be ruled out by the CDE)

To determine the coefficients in Eq.~(\ref{eq:R}) we shall fit the
experimental values of charge radii (full references are given
in~\cite{duf94}), reduced to point radii through the standard
prescription $\rho_{\pi}^{exp} =[\langle
r^2_{\pi}\rangle_{charge}^{exp} -0.64]^{1/2}$ (used in~\cite{trz01},
full form in~\cite{cha97}).

A fit is made to 82 experimental values for nuclei with $N$ or $Z=6,\,
14,\, 28,\, 50,\, 82,\, 126$ (the classical closures which we call EI,
for extruder-intruder), for which shell effects can be assumed to be
minimal. The results are altogether remarkable: {\em for $0.4\le
  \zeta\le 1.2$ the root mean square deviations {\em (rmsd)} are
  below} 10 mf.  We find nearly common parameters $\rho_0=0.943(2)$,
$g=1.04(3)$, for different ($\upsilon\, ,\zeta$) pairs, given in
Table~I for $0.6\le \zeta\le 1.0$. 

To decide which is the favored ($\upsilon\, ,\zeta$), we compare 63
experimental~\cite{aud95} displacements,
CDE=BE($Z_>,N_<$)-BE($Z_<,N_>$)---where $Z_>=N_>=\max
(Z,N)$\footnote{We shall use throughout $Z_>$ ($Z_<$) for ``the
  proton (neutron) excess
  mirror partner''.}---with
calculations. (We have followed~\cite{bro00b} where, out of 78 values,
those involving proton-unbound states are discarded). In the next
paragraphs we define one by one the ingredients and then examine their
effect column by column in Table~I.

The Coulomb potential $V_C$ will be written in an oscillator
representation.  To conform to standard use we introduce
$R_{\pi}=(5/3)^{1/2}\rho_{\pi}$. One of the many advantages of the
oscillator basis is that one can separate trivially an adimensional
two-body $\widehat{\left(1/r\right)}$ operator through the first
equality below
\begin{equation}
  \label{eq:vc0}
  V_C=e^2\sqrt{\frac{M\omega_{\pi}}{\hbar}}\;\widehat{\left(\frac{1}{r}\right)}
=\frac{1.934}{R_{\pi}}Z^{1/6}\;\widehat{\left(\frac{1}{r}\right)}\,
{\rm MeV}, 
\end{equation}
The second equality provides the connection between CDE and radii. It
follows from a famous estimate~\cite[Eq.~(2-157)]{BM69}, which we
apply separately to neutrons and protons (MeV units):
\begin{equation}
  \label{eq:hw}
\frac{\hbar\omega_{\pi}}{(2Z)^{1/3}}=
\frac{35.59}{\langle r_{\pi}^2\rangle}; \hspace{.2cm}
\frac{\hbar\omega_{\nu}}{(2N)^{1/3}}=\frac{35.59}{\langle r_{\nu}^2\rangle}.
\end{equation}

The  $\widehat{\left(1/r\right)}$ operator will be evaluated in detail
later, and it will be shown that Eq.(\ref{eq:vc0}) leads to a
``smooth average of the diagonal monopole part'',
\begin{equation}
\label{eq:vcc}
\langle V_{Cm}^d\rangle \approx\frac{.864\, (Z(Z-1)-
  Z)}{R_{\pi}^{c}}\, {\rm MeV}, 
\end{equation}
identical---within a small exchange term---to the classical expression
for the charged sphere. ($R_{\pi}^{c}$ is the charge, not the point
radius, as explained after Eq.~(\ref{eq:vc}).)

The column labelled $\langle V_{Cm}^d\rangle$ in Table~I gives the rms
deviations when only this term is included. The preferred value is at
$\zeta\le 0.6$.

The Bethe Weizs\"acker form for the charge symmetry breaking (CSB) 
potentials is 
\begin{equation}
  \label{eq:vb}
  V_B=-\frac{t}{2}\left [B_v\, -B_s\, (\frac{A_0}{A})^{1/3}\right ]\, {\rm keV}.
\end{equation}
The parameters $B_v$ and $B_s$ {\em cannot} be chosen arbitrarily.
They are not known experimentally, but enormous theoretical work has
been devoted to CSB potentials, extensively described in
Ref.~\cite{mac01}, from which we take the $V_B$ contribution to be
$\approx 100$ keV around $A=16$, and $\approx 300$ keV in nuclear
matter.  Accordingly, we set $A_0=16$ and define a standard $V_B^{st}$
at $B_v=300$ keV and $B_s=200$ keV.  Calling $V^{st}=V_{Cm}^d+V_B^{st}$, the
corresponding column shows that the preferred value moves to
$\zeta=0.8$, with a slight gain in rmsd.

\begin{table}
\caption{{\label{tab:mde}}Root mean square deviations from observed
  CDE for three sets of coefficients in Eq.~(\ref{eq:vb}) with
 $A_0=16$. $V_C$ from Eq.~(\ref{eq:vc}).  All energies in keV. See text.} 
\begin{ruledtabular}
\begin{tabular}{ccccccc}
&&{$V_{Cm}^d$}&{$V^{st}$}&
\multicolumn{3}{c}{$V_{Cm}^d+V_B$}\\  
\hline
$\zeta$&$\upsilon$&rmsd&rmsd&$B_v$&$B_s$&rmsd\ \\
\hline  
1.0&0.51&{\bf 540}&{\bf 272}&540&350&{\bf 127}\\  
0.9&0.57&{\bf 434}&{\bf 184}&431&281&{\bf 127}\\
0.8&0.62&{\bf 320}&{\bf 127}&306&199&{\bf 127}\\
0.7&0.68&{\bf 213}&{\bf 170}&181&119&{\bf 127}\\
0.6&0.73&{\bf 136}&{\bf 268}&\ 56&\ 39&{\bf 127}\\  
\end{tabular} 
\end{ruledtabular}
\end{table}
 
In the last three columns $B_v$ and $ B_s$ are allowed to vary
simultaneously to give an idea of the range of plausible $V_B$
parameters. (Note the advantage of introducing $A_0$ in
Eq.~(\ref{eq:vb}): $B_v-B_s$ equals the contribution at $A=16$.) The
constancy of the rmsd indicates the absolute need of constraints on
$V_B$. Conversely, the rapid variation of $V_B$ constrains the skin to
a narrow range of plausible values for $B_v$ and $B_s$. We
propose\footnote{The error bars are probably too large. R. Machleidt
  considers $B_v=300$ keV to be an upper limit (private
  communication).}  $\Delta_{r_{\nu\pi}}= 0.80(5)\, t/A$, (nearly)
compatible with Eq.~(\ref{eq:s1})~\footnote{It is often argued that
  measures are model dependent and sensitive to an ``interaction''
  radius rather than to $\Delta_{r_{\nu\pi}}$ proper. Our answer to
  this objection is threefold. A) we are unaware of estimates
  indicating that uncertainties are larger than the---large---quoted
  errors. B) If anything, the effect would lead to smaller
  $\Delta_{r_{\nu\pi}}$, as our calculations do indeed. C) We make no
  use whatsoever of the experimental results. The reader may decide
  for himself whether he wants to compare them with our estimate or
  not.} and (perfectly) compatible with the estimate for $^{208}$Pb
in~\cite{bro00}.

The results depend on the validity of Eq.~(\ref{eq:R}) which was
derived assuming isospin purity for the wave functions. The assumption
is likely to hold near stability, but approaching the proton drip line
it becomes more questionable. This is why we followed
Ref.~\cite{bro00b} in rejecting data from 15 pairs with proton-unbound
$Z_>$. If they are reincorporated, the numbers are telling: the
preferred $\zeta$ moves slightly up at 0.85 but the rmsd doubles at
254 keV. Some of the rejected pairs do fairly well, but for eight of
them the error is above 500 keV. (In the restricted sample no error is
larger than 300 keV). These findings are consistent with good $T$ near
stability and deteriorating away from it. At stability, the $T\approx
0$ nuclei raise the problem of a negative skin, small for
$A\lessapprox 60$ and quickly dissappearing as $T$ increases (see for
example~\cite{col95}).  Our calculation should detect it as a slight
deterioration for $T=1/2$ with respect to $T=3/2$ or higher. If
anything, what is seen is the opposite. This observation does not
contradict the results of Ref.~\cite{col95}; it simply suggests that
the hitherto neglected shell effects are stronger than those induced by 
$T$-violation.

\vspace{.3cm}
 
Let us examine now the smooth forms we have introduced. The radii
enter only through the oscillator frequencies in Eq.~(\ref{eq:hw}),
which define the basis in which we should work. For magic nuclei the
basis reduces to a single state, and the observed radii can be assumed
to be given by $\rho_{\pi}$. When moving away from closed shells,
configuration mixing will increase the size of the system: moderately
for nearly spherical nuclei, and radically for deformed ones. The
crucial point is that {\em $\rho_{\pi}$ is not supposed to produce good
radii, but a good basis}. To check whether this is the case we propose
the following test: Try to fit all the available data through a shell
corrected radius $\rho_{\pi}^{sc}=\rho_{\pi}+ {\cal D}$, fixing
$\rho_{\pi}$ to the values we have obtained. The test is passed if the
fit is good and nothing is gained by re-fitting $\rho_{\pi}$.

\begin{figure}[ht]
  \begin{center}
    \leavevmode
    \epsfig{file=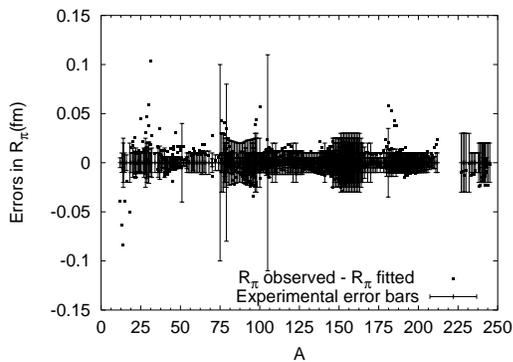,width=7cm}
    \caption {Comparison between experimental errors and differences
      between observed and calculated radii
      $R_{\pi}=(5/3)^{1/2}\rho_{\pi}$ ($\zeta=0.8$, undistinguishable
      from other choices.)}
    \label{fig:ray}
  \end{center}
\end{figure}

The definition of ${\cal D}$ is suggested by previous work on shell
corrections~\cite{duf94,zuk94,dz95}. We have

\begin{equation}
\rho_{\pi}^{sc}=\rho_{\pi}+ {\cal D}, \hspace{.3cm}
{\cal D}=\lambda S_{\pi}S_{\nu}+\mu Q_{\pi}Q_{\nu}.
\end{equation}
\noindent
$S$ stands for spherical, $Q$ for deformed. The correction ${\cal D}$
is a functional of the orbital occupancies in the EI
(extruder-intruder) valence spaces. They consist in the orbits of
harmonic oscillator shell $p$, except the largest (extruder), plus the
largest (intruder) orbit from shell $p+1$.  For protons, say, the EI
degeneracy is $D_{\pi}=(p_{\pi}+1)(p_{\pi}+2)+2$ (e.g., $p_{\pi}=3$
between $Z=28$ and $Z=50$). The degeneracy of the non intruder orbits
is $D_{r\pi}=p_{\pi}(p_{\pi}+1)$. Now call $z$ the number of valence
protons, and define $S_{\pi}=z(D_{\pi}-z)/D_{\pi}^2$,
$Q_{\pi}=z(D_{r\pi}-z)/D_{\pi}^2$; and similarly for neutrons.  By
construction, ${\cal D}$ vanishes at the EI closures.

Fits of $\rho_{\pi}^{sc}$ to 634 experimental values keeping the
parameters previously obtained for $\rho_{\pi}$---variation leaves
them unchanged---lead again to remarkable results: with
$\lambda=5.6(2)$, $\mu=23(1)$ we obtain rmsd $\le$ 11 mf in all cases.
Fig.~\ref{fig:ray} gives an idea of the quality of the results: most
of the calculated points fall within (or very close to) the
experimental error bars. The exceptions are the light nuclei (where
halo orbits are important), and the region around the light Pt
isotopes known for shape coexistence in the ground states. The test is
passed: $\rho_{\pi}$ makes sense. It should be noted that a fit of
Eq.~(\ref{eq:R}) to all nuclei makes little sense.

The Bethe Weisz\"aker form for $V_B$ is obvious.

The Coulomb potential is a different matter. The quality and
credibility of our approach rests on an {\em exact} treatment of
$V_{Cm}^d$.

Following Ref.~\cite{dz96} we separate the Hamiltonian $H=H_m+H_M$.
The monopole part $H_m$ contains all terms in scalar products of
fermion operators ($a^+_r\cdot a_s$) for sub-shells $r$ and $s$. We
consider the {\em diagonal} part involving the number operators
$m_r=a^+_r\cdot a_r$. The non-diagonal one---responsible for isospin
impurities---will be ignored, as will the multipole $H_M$ which
contributes to shell effects~\cite{dz96,zuk94,dz99,zuk02}.

Calling $V_{iki'k'}$ the matrix elements of $\widehat{1/r}$,
the diagonal monopole part is
\begin{equation}
\label{eq:mono}
\widehat{\left(\frac{1}{r}\right)}_m^d=\sum_{i\leq k}\frac{z_i(z_k-\delta_{ik})}{1+\delta_{ik}}V_{ik},\;\; 
V_{ik}=\frac {\sum_{J}V_{ikik}^J[J]}{\sum_{J}[J]},
\end{equation}
$[J]=2J+1$, $z_k=$ number of protons in orbit $k$, where $k\equiv plj$
stands for the quantum numbers specifying a given harmonic oscillator
(ho) orbit ($p$ is the principal quantum number).  The sum over the
first $\kappa$ major shells containing $\tau$ orbits can be reduced to
a sum of factorable terms by diagonalizing the matrix
$\frac{1}{2}\{V_{ik}\}$. This technique is extensively described
in~\cite{dz96}. As the highest eigenvalue $E_{\tau}$ overwhelms all
others, when combined with the corresponding eigenvector $U_k$ and the
single particle counter term in Eq.~(\ref{eq:mono}) it determines the
largest contribution by far ($z_k$ is the number operator for orbit $k$)
\begin{equation}
  \label{eq:vcfac}
\widehat{\left(\frac{1}{r}\right)}_m^d=\frac{1}{2}
\left[ E_{\tau}\left(\sum_k z_k U_{k}\right)^2-\sum_k z_k\, V_{kk}\right],
\end{equation}
{\em which amounts to a basically exact representation in Fock space}:
$E_{\tau}$ is 30 times bigger than the second biggest eigenvalue, and
100 times bigger than the third. Furthermore---as we shall see--$U_k$
leads to an extremely coherent operator, while the other eigenvectors
do not. We have treated the $\kappa=8$, $\tau=36$ case, but the
results are independent of the number of orbits. 

\begin{figure}[ht]
    \begin{center}
      \leavevmode
      \epsfig{file=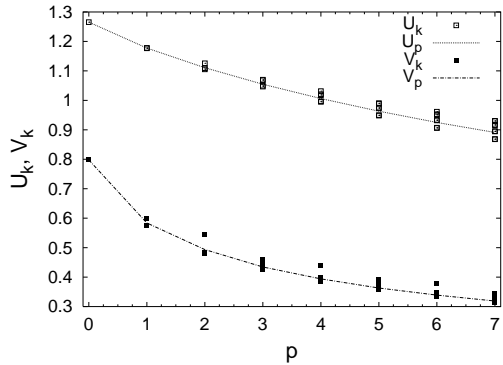,width=7cm}
   \caption {The terms in Eq.~(\ref{eq:vcfac}). See text} 
      \label{fig:dz}
    \end{center}
\end{figure}

Fig.~\ref{fig:dz} shows the forms of $U_k$ and $V_{kk}\equiv V_k$. To
deal with quantities of order 1, we have rescaled $U_k\Longrightarrow
\sqrt(\tau)\, U_k$ and accordingly $E_{\tau}\Longrightarrow \tau\,
E_{\tau}=.383$.  The $(2j+1)$-weighted averages over $j$-orbits within
a major shell $p$ are also shown ($U_p$ and $V_p$). By referring the
$l(l+1)$ term to its centroid we have to good approximation
\begin{equation} 
\label{eq:ll}
U_k=U_p- 0.01\, [l(l+1)-p(p+3)/2]/(p+3/2) 
\end{equation}
For $V_k$ the $j$ dependence is more complicated. To leading order in
a $(p+x)^{-y}$ expansion the averages are $U_p\approx
1.522(p+1.4)^{-1/4}$, and $V_p\approx 0.93(p+3/2)^{-1/2}$.

To extract a smooth form, we fit separately the two terms of
Eq.~(\ref{eq:vcfac}) at the $p$ closures (i.e. replacing the sum over
$k$ by a sum over $p$, with $z_k\equiv (p+1)(p+2)=D_p$, $U_k\equiv
U_p$, and $V_k\equiv V_p$) to forms $a(Z-b)^c$. Exact values for
$U_p$, and $V_p$ are kept. Then we insert the factor in
Eq.~(\ref{eq:vc0}), to obtain (MeV units)

\begin{equation}
    \label{eq:vc}
\langle V_{Cm}^d\rangle \approx \frac{.858\,
  Z^{1/6}\left[(Z-1/2)^{2-1/6}-Z^{1-1/6}\right]}{R_{\pi}}.
\end{equation}
\noindent
{\em The numerical uncertainties of the fit ($\approx$ 0.1\%) allow
  the choice of the round numbers shown}. This is the form used in the
calculations. For clarity, in Eq.~(\ref{eq:vcc}) it has been very
slightly (and innocuously) simplified and the point radius parameter
$R_0=\sqrt{5/3}\rho_0=1.219$ replaced by its charged value
$R_0^c=1.226$, so as to change the overall coefficient, which becomes
the one for the classical charged sphere: $3e^2Z^2/5R^c\approx
0.864Z^2/R^c\approx0.7Z^2/A^{1/3}$.  A pleasing result.

\begin{figure}[ht]
  \begin{center}
    \leavevmode
    \epsfig{file=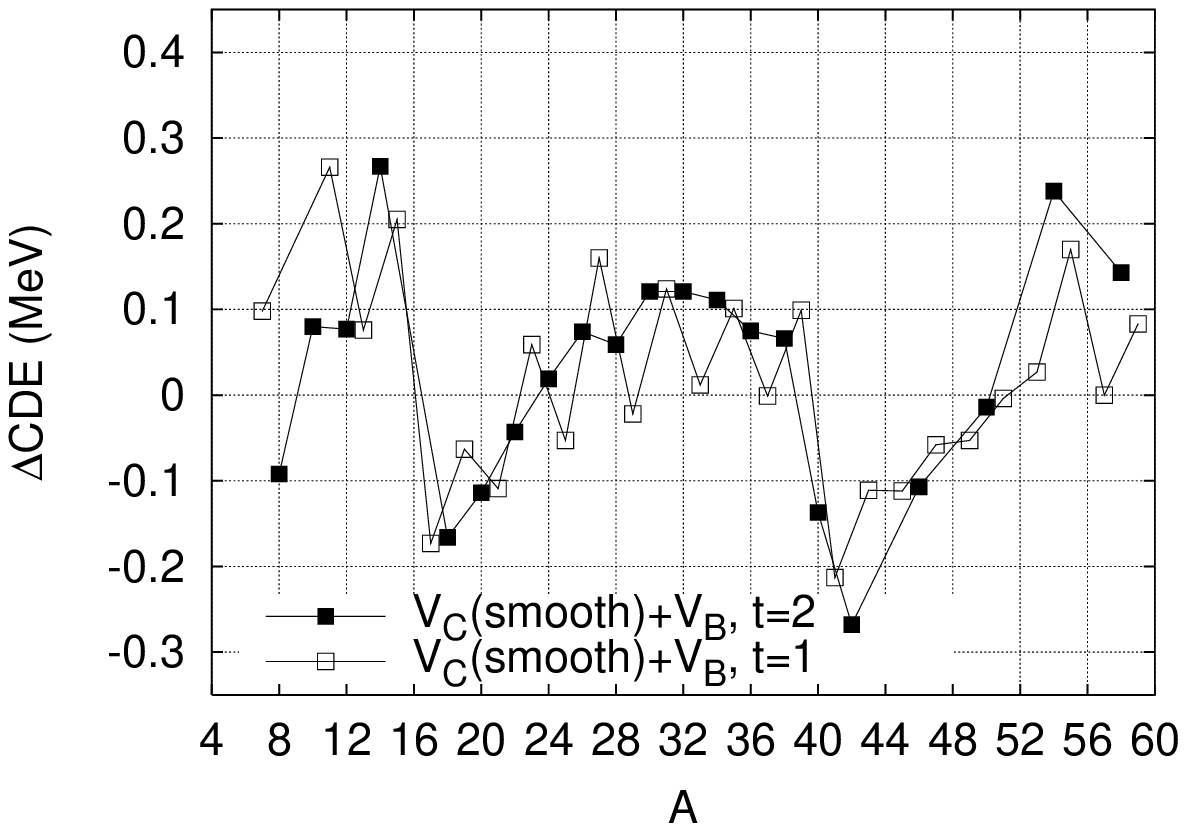,width=7cm}
    \epsfig{file=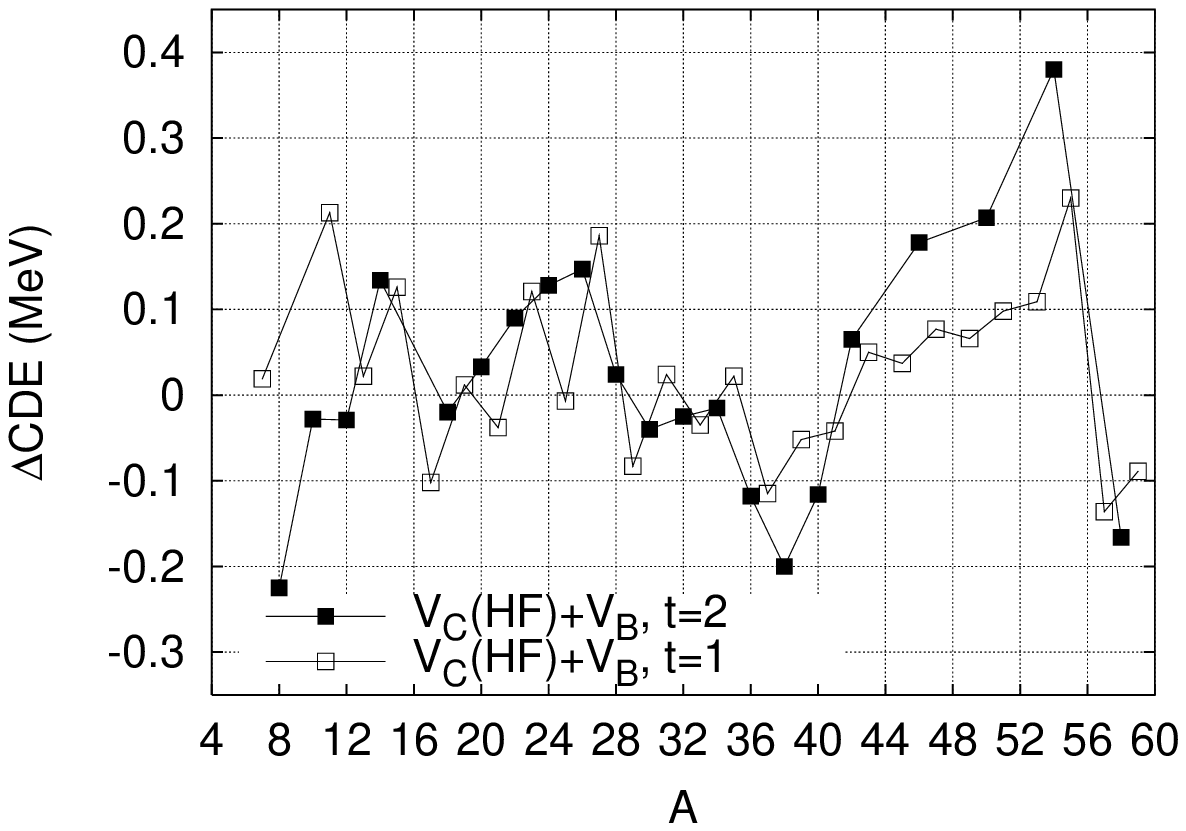,width=7cm}
    \caption{First panel: Experimental$-$calculated CDE for $t=1$ and
      2. witht smooth occupancies ($\zeta=0.8$). Second
      panel: As above for Hartree Fock occupancies.}
    \label{fig:mde2}
  \end{center}
\end{figure}

The form of averaging we have used eliminates all shell effects, which
show cleanly in the first panel of Fig.~\ref{fig:mde2} as the
difference between experimental and calculated values. For clarity
only $t=1$ and 2 cases are kept. They are sufficient to indicate the
secular nature of the deviations.
 
Smooth filling can be replaced by {\em ordely} filling, which, using
exact values for $U_k$ and $V_k$ in Eq.~(\ref{eq:vcfac}), should be
close to a Hartree Fock (HF) result, since Coulomb matrix elements are
quite unsensitive to details of the single particle wavefunctions,
except in the case of halo orbits. 

The difference between smooth and HF approximations is mostly due to
the $l(l+1)$ term in Eq.~(\ref{eq:ll}), whose influence can be
understood by noting that above a closed shell with $Z=Z_{cs}$ it
produces a single particle field
\begin{equation}
  \label{eq:epsc}
\varepsilon_{Cl}=\frac{-4.5\, Z_{cs}^{13/12}\,
  [2l(l+1)-p(p+3)]}{A^{1/3}(p+3/2)}\, {\rm keV},  
\end{equation}
obtained from Eq.~(\ref{eq:vc0}) and Eq.~(\ref{eq:vcfac}), using 
$E_{\tau}=0.383$, and  the coefficient in $U_p\approx
1.522(p+1.4)^{-1/4}$, mentioned after Eq.~(\ref{eq:vcfac}).

The HF displacements in the second panel of Fig.~\ref{fig:mde2}
indicate that the $l(l+1)$ effect leads to a dramatic improvement for
the $A=15$-17 and $A=39$-41 pairs, the two test cases most extensively
studied in the literature (\cite{nol69,mac01,agr01} and references
therein).  This gain is upset by a loss of regularity in the
shell-effect patterns and the (apparent) need for a too strong $V_B$.
If we keep $V_B^{st}$ (and $\zeta=0.8$) the rmsd becomes large (330
keV). The figure corresponds to a fit (rmsd=150 keV) that lead to a
huge change in $V_B$, at $B_v-B_s=425-180=245$ keV.

The situation is in all respects similar to that in
Ref.~\cite{bro00b}, where a full HF description with the SKXcsb force
also demands a too large $V_B$ contribution (355 keV at $A=17$).  The
agreement between~\cite[bottom Fig. 2c]{bro00b} and the second panel
of Fig.~\ref{fig:mde2} is quite good, to within an overall shift of
some 100 keV for the latter~\footnote{In~\cite[bottom Fig. 2c]{bro00b}
  crosses and squares should be interchanged. Then it only takes good
  eyes to check the agreement.}. A convincing indication that our
approximation to a full HF result is sound. However, it casts doubts on
the very use of the HF approach:

\noindent
Near closed shells it is likely to be valid, and the improvement it
brings about is very beneficial. Elsewhere, HF produces misleading
patterns through shell effects that should be treated in conjunction
with those of multipole and CSB origin. Unless this is done, it is
clearly better to keep the smooth approximation.

To conclude, we examine what can be said about the Nolen Schiffer
anomaly (NSA), which we define as a systematic failure to obtain good
CDE without {\it ad-hoc} adjustable parameters. The historical origin
of the problem is clear from column $V_{Cm}^d$ in Table I, showing
that the Coulomb potential leads to large discrepancies for acceptable
values of the neutron skin~\footnote{This statement of the problem is
  compact and accurate but non-standard. In the late 1970's some
  experiments appeared to be consistent with vanishing neutron skin,
  and it was pointed out that under this assumption the NSA would
  disappear~\cite{shl77}. After this proposal turned out to be
  unrealistic, with few exceptions (such as~\cite{gom83}) little
  emphasis was put on the strong correlation between CDE and neutron
  skin.}. It is also clear that the $V_B$ potentials though small (of
the order of the Coulomb exchange), are crucial in moving the skin to
a reasonable range. As a consequence, the remaining discrepancies are
the shell effects in the first panel of Fig.~\ref{fig:mde2}.

As we have seen, for the classical test cases---at $A=15$, 17 and
$A=39$, 41---the $l(l+1)$ term reduces the errors (to some 100 keV for
the former and 50 keV for the latter). To achieve greater precision,
good quality shell model calculations are necessary, a difficult task,
not undertaken so far, at least for the CDE. Therefore, at present, we
cannot decide whether there subsists an anomaly or not.

For nuclei with several particles in a major shell, high quality
configuration mixing is possible, and the task has been undertaken,
not for the CDE, but for the differences of excitation energies
between mirror yrast bands (CED, better called MED)~\cite{zuk02}:
Shell effects play a major role, and the CSB contribution is at least
as important as the Coulomb one. The results achieve an accuracy of
$\approx 10$ keV on differences of up to 100 keV. This is the order of
magnitude of the discrepancies we want to correct. Furthermore, the
CED are unlikely to demand better control of the wavefunctions and the
interactions than the MED do. As a consequence---though we cannot
decide whether there remains a problem until the CED calculations are
done---a fairly safe bet is that the Nolen Schiffer anomaly will
disappear.

\vspace{.6cm}

We thank J. Bartel for his patient coaching on Skyrme calculations, G.
Mart\'{\i}nez for his help, and R. Machtleidt, M. Horth-Jensen and P.
Vogel for their comments.

\end{document}